\documentclass[
reprint,
amsmath,
 pre
]{revtex4-1}

\usepackage{graphicx}
\usepackage{float}
\usepackage{dcolumn}
\usepackage{bm}
\usepackage{upgreek}

\usepackage{lineno}
\begin{document}

\title{Identifying Diffraction-limited Single-Molecule Nanofluidic Motions}






\author{Siddharth Ghosh$^\dagger$}
\email{sg915@cam.ac.uk}
\altaffiliation{also affiliated to Single-Molecule Optics Group, Huygens Laboratory, Leiden Institute of Physics, Leiden University, The Netherlands. St John's College, University of Cambridge, Cambridge, UK.}
\affiliation{Department of Applied Mathematics and Theoretical Physics, University of Cambridge, Cambridge, UK.\\
Yusuf Hamied Department of Chemistry, University of Cambridge, Cambridge, UK.\\
Maxwell Centre, Cavendish Laboratory, University of Cambridge, Cambridge, UK.\\
 Open Academic Research UK CIC, Cambridge, UK and Open Academic Research Council, Kolkata, India.}

\begin{abstract}
    Single-molecule motions in the nanofluidic domain are extremely difficult to characterise because of various complex physical and physicochemical interactions.  
    We present a method for quasi-one-dimensional sub-diffraction-limited nanofluidic motions of fluorescent single molecules using the Feynman-Enderlein path integral approach.   
    This theory was validated using the Monte Carlo simulation to provide fundamental understandings of single-molecule nanofluidic flow and diffusion in liquid.  
    The distribution of single-molecule burst size can be precise enough to detect molecular interaction.  
    The realisation of this theoretical study considers several fundamental aspects of single-molecule nanofluidics, such as electrodynamics, photophysics, and multi-molecular events/molecular shot noise.
    We study  {molecules within (an order of magnitude of) realistic lengthscale for organic molecules, biomolecules, and nanoparticles where  {1.127 nm} and  {11.27 nm} hydrodynamic radii of molecules were driven by a wide range of flow velocities ranging from $0.01~\upmu$m/s to $10~\upmu$m/s}.  
    It is the first study to report distinctly different velocity-dependent nanofluidic regimes.  
\end{abstract}

\maketitle

\section*{Introduction}
Statistical mechanics at nanometric length scale can unravel many questions related to life on earth and reveal unfamiliar physics \cite{schrodinger2012life, loose2008spatial, litschel2018freeze, golestanian2015enhanced, illien2017exothermicity, ghosh2016ferroelectric, ghosh2016atomic}.  
Tracking of single molecules \cite{enderlein2000positional} has been a centre of this field \cite{Cohen2006, CohenPNAS2006}.
The state-of-the-art nanoengineering provides sufficient precision to handle individual single molecules by overcoming diffusion induced slip from detection volume and heterogeneous electrodynamics-induced artefacts in single-molecule fluorescence signals. 
Nanofluidic motions or confined flow and diffusion environment plays a crucial role where two spatial dimensions are in the order of the mean free path of single molecule's diffusion or tighter spatial confinement than the wavelength of light being used for single-molecule detection \cite{ghosh2020single}. 
Such environments are molecular motion inside tunnelling nanotube or membrane nanotube \cite{ranzinger2011nanotube}, nanopores, single molecule sequencing - linearise DNA in nanochannel array \cite{jeffet2016super}, zeolite-catalysis \cite{zoran2018reversible},
and mass transfer in carbon nanotubes \cite{Noy2006, Noy2017}. 
 In dynamic single-molecule experiments \cite{lu1998single, CohenPNAS2006, zijlstra2012optical, baaske2014single, lesoine2012nanochannel, ghosh2020single}, one of the hardest problems is to identify if the signals are purely single-molecule events due to the convoluted complexity associated with electrodynamic interactions/Purcell effect \cite{purcell1995spontaneous, enderlein2000theoretical, karedla2015simultaneous, Tegenfeldt2010}, photophysics \cite{eggeling2001data, orrit2009langmuir}, electrostatic effect \cite{debye1923theorie, french2010long}, multi-molecular interactions \cite{Joerg1998}, and confined diffusion \cite{malek2003knudsen, li2019diffusion}.
Diffraction-limited detection of these single-molecule nanofluidic motions possess two problems -- 
(a) how do we know that a single molecule is flowing or crawling (non-Brownian interfacial diffusion \cite{wang2020non}) in the detection volume? 
(b) How do we know that more than one molecule is present in the detection volume?
Photon antibunching \cite{basche1992photon} and step-wise photobleaching \cite{Ghosh2014} are two widely used methods to resolve this spatio-temporal problem. 
Those two strategies are not useful for dynamic systems since photon statistics is limited by the number of photons.  
 {However, in time-dependent studies, burst sizes (the duration of single or multiple photon bursts) are  efficient approach to quantify fundamental interactions, if we are inspired from the field of astrophysical studies of Gamma-ray bursts\cite{piran2005physics}.}

\begin{figure*}[t!]
    \centering
    \includegraphics[width=\textwidth]{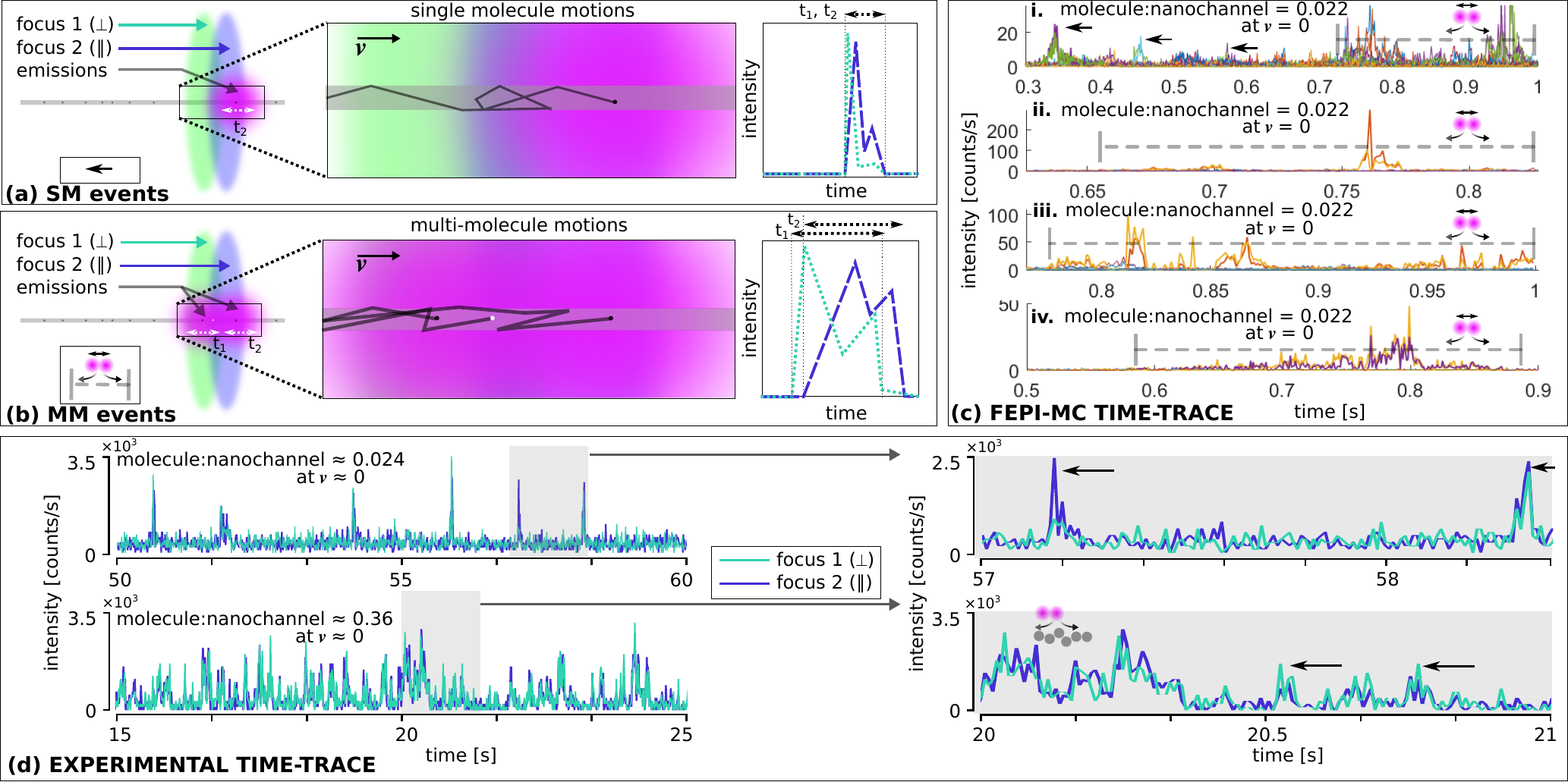}
    \vspace{-3mm}
    \caption{\textbf{Single-molecule and multi-molecule events inside a { nanofludic} channel that is narrower than the diffraction limit.} (a) { Nanofluidic channels placed inside two confocal foci -- green and purple colour represents orthogally polarised light from each other.} The black dots represent single fluorophore molecules; in the detection volume photon, emission is represented with magenta circle. (b) Multi-molecule (MM) event with more than one single molecules or crawling/dragging single-molecules may result in photobleached/photo-inactive molecules  -- white dot.  (c) FEPI-MC simulated two-foci SM events (i and ii) and MM events (iii and iv) of  {1.127 nm} molecules in 50 nm nanofluidic channel  {(ratio of these diameters is 0.022} at zero velocity, exemplary SM events are shown using the arrows.
     (d) Experimental two-foci single-molecule fluorescence intensities of  {Atto-488 in 50 nm nanochannel (molecule:nanochannel $\approx$ 0.024  approximately zero velocities)} and 48bp DNAs flowing through a 30 nm nanochannel  {(molecule:nanochannel $\approx$ 0.36 at approximately zero velocities)}; details of grey regions are shown in the left panel. 
     }
     
    \label{fig:3}
    \vspace{-4mm}
\end{figure*}

In this report, we demonstrate a method to predict and model the characteristics of single-molecule  {fluorescence} bursts and their spatio-temporal distributions inside a nanofluidic environment.
Transiting single molecules through { {diffraction-limited confocal detection volumes produces photon bursts, which can be time traced and quantified as burst size. 
Time duration of each single-molecule bursts in transient spectroscopy provide a statistics}} of burst size distribution { {(BSD)}} \cite{Joerg1998} or burst variant analysis with dynamic heterogeneity and static heterogeneity \cite{torella2011identifying, eggeling2001data} provides useful information to distinguish between molecular shot noise \cite{chen1996single}, multi-molecule events, and single-molecule events. 
A general framework is always sought after that is independent of any initial assumption of the experimental condition.  
In section II, we used a general and first-principles method of direct mathematical approach of path integral formalism to consider diffusion and photo-physical processes. 
Thus, it is a perfect initialisation to formulate a high-throughput Monte Carlo simulation of this kind of problem.
In section III, the Monte Carlo simulation of the path integral formalism { {(or FEPI-MC/Feynman-Enderlein path integral Monte Carlo)}} is presented for two different sizes of single molecules,  {1.127 nm} and  {11.27 nm}  {since majority of the small organic molecules, biomolecules, and nanoparticles are within this range}. 
 {We have performed the simulation for two overlapping confocal foci as used in two-foci correlation spectroscopy \cite{Joerg2007, chiantia2006combined}.}
 {The advantage of using two foci is to extract the velocity of the flowing single molecules, which is difficult to obtain in a single focus \cite{dittrich2002spatial, nienhaus2013studying}}. 
In section III, we have shown the burst-size distribution of the single-molecule nanofluidic motion using FEPI-MC simulation and identified the new nanofluidic regimes.
 {Experimental single-molecule fluorescence bursts inside all-silica nanofluidic channels are shown for qualitative comparison with the numerical findings}.

\section*{Computing Burst Size Distributions} 
\textit{Model:} 
We used the Feynman path integral \cite{feynman2005feynman} to model single-molecule BSD using the burst searching theory developed by Enderlein et al.~\cite{enderlein1997statistics, Joerg1998}. 
Hence, we termed the modified path integral for single-molecule fluorescence as Feynman-Enderlein path integral  {(FEPI)}. 
We looked into the temporal probability of the existence of an unbleached single-molecule while passing through a confocal laser beam (Supplementary Figure 1). 
The existence probability decays exponentially after detecting a photon when no further photon is detected. 
Once the next photon gets detected instantaneously the probability goes to a larger value due to the possibility of a photon being emitted from the molecule instead of the background. 
The non-vanishing probability of background photon keeps the existence probability always less than unity. 
This time evolution of the existence probability is an important model for single-molecule burst identification. 
It is unavoidable to ignore the photobleaching effect from the expression of the existence probability within a given time interval in the case of single-molecule nanofluidics. 
Derivation for the photobleaching contribution can be found in \cite{enderlein1997statistics}. 

For pure single-molecule transits, let us consider that at $\tau = 0$ when the molecule with trajectory $\bm{r}(\tau)$ started flowing outside the confocal volume, and $\tau = T$ was large enough to consider that the molecule has crossed the detection volume.  
The existence probability of single-molecule BSD is given by $P_1(N)$ where 1 stands for `single-molecule crossing'. 
In the model, we considered one-step photobleaching and negligible triplet-state (long triplet state can be handled as well). 
A simple Poisson distribution describes the photon detection statistics of these single molecules' sub-ensemble as
\begin{equation}
    X_f = \int_0^\tau X_f[\bm{r}(\tau)] \text{d}\tau
\end{equation}
\noindent where within the time interval $\delta \tau$ the molecule at position $\bm{r}$ gives rise to the probability of detecting a photon as $X_f(\bm{r}) \delta \tau$. The probability of detecting $N$ photons during the time interval $\{0,T\}$ will be superposition of these Poisson distributions as (equation 2),
\begin{equation}
    P_1 (N) = \int_0^\infty \frac{X_f^N}{N!} \exp (-X_f) \cdot P[X_f] \text{d}X_f
\end{equation}
\noindent where $ P[X_f]$ is the distribution function in a form of a path integral, which runs over all possible photobleaching times and molecular trajectories.

\begin{multline}
    P[X_f] = \int_0^T \int D\bm{r}(\tau) \delta\Bigg[X_f - \int_0^{\tau_{bl}} X_f[\bm{r}(\tau)]\Bigg] X_{bl}[\bm{r}(\tau_{bl})]\text{d}\tau_{bl} \times \\
    \exp\Bigg[-\int_0^{\tau_{bl}} \Bigg(\frac{[\bm{r}(\tau) - \bm{v}]^2}{4D} +
    X_{bl}[\bm{r}(\tau)]) \text{d}\tau\Bigg)\Bigg]p_0[\bm{r}_0] + \\
    \int D\bm{r}(\tau)\delta\Bigg[X_f - 
    \int^T_0 X_f[\bm{r}(\tau)] \text{d}\tau\Bigg] \times \\
     \exp \Bigg[-\int^T_0 \Bigg(\frac{[\bm{r}(\tau) - \bm{v}]^2}{4D} + 
    X_{bl}[\bm{r}(\tau)]) \text{d}\tau\Bigg)\Bigg] p_0[\bm{r}_0]
\end{multline}

\noindent The $\delta[f(\bm{r})]$ is $\delta$-function functional, $p_0(\bm{r_0})\delta\bm{r}$ is the probability of finding a molecule in the small volume $\delta\bm{r}$ at position $\bm{r}_0$ at time $t = 0$, $\delta \tau X_{bl}\bm{r}$ is the probability of photobleaching the molecule within $\delta\tau$  at position $\bm{r}$, $D$ is the diffusion constant of the molecule, and $\bm{v}$ is the flow velocity.  
The path integration $\int D\bm r(\tau)$ in equation 3, runs over all possible paths starting from a random point $\bm{r}_0$ and random endpoint. 
The first and second terms in equation 3 represent the contribution of molecules that photobleach and do not photobleach, respectively where the $X_f(\bm{r}) = \eta(\bm{r}) \Phi_f \sigma I (\bm{r})$ and $    X_{bl}(\bm{r}) = \Phi_{bl} \sigma I (\bm{r})$. 
Here, $I(\bm{r})$ is the spatial dependence of intensity of the confocal laser beam (photons/area/time), $\eta(\bm{r})$ is the collection and detection efficiency of optics and electronics, $\Phi_f$ and $\Phi_{bl}$ are the fluorescence and photobleaching quantum yields, and $\sigma$ is the absorption cross-section. 
Since the intensity of the excitation beam is low, no optical saturation effect is present.
The all-silica nanochannels show negligible background and a steady background with respect to observation time \cite{ghosh2020single}. 
So, we neglect the background effects. 
The analytical expression of $P_1(N)$ was not achievable -- equation 3 was run over an infinite number of paths. 
Thus, it was fundamentally difficult to find a solution. 
Monte-Carlo sampling was done to calculate the $P_1(N)$ where random paths were chosen and the remaining integration was performed numerically. 
A large number of random sampling over different paths led to a sufficiently precise approximation of $P_1(N)$.

Multi-molecule events or close succession of more than one molecule flows and lead to a single photon burst always have a nonzero probability. 
This causes additional peaks with an extra-temporal broadening of the photon burst at larger burst sizes than single molecule events. 
We use a weighting factor $w_k$. 
It is proportional to the probability that $n$ molecules pass through the confocal volume, which are separated by less than the mean transit time leading to a single unresolved burst. 
Due to these $n$-molecule bursts, the BSD is the convolution of BSD for all $(n-1)$-molecule bursts along with the BSD of the pure single-molecule burst. 
So, BSD for two-molecule is given by
\begin{equation}
    P_2(N) = \sum_{N} P_1(N_1) P_1(N_2)
    \vspace{-2mm}
\end{equation}
\noindent where $N = N_1+N_2$, and BSD for $n>2$-molecule is given by
\begin{equation}
    P_n(N) = \sum_{N_*} P_1(N_1) P_1(N2) ... P_1(N_n) 
    = \sum_N P_1(N_1)P_{n-1}(N_2)
\end{equation}
\noindent where $N_* =  N_1+N_2+...+N_n$.
If $\tau_s$ is the minimum time between molecules leading to an unresolved prolonged burst, $\tau_m$ is the initial interval time during delivery of the molecules from the reservoir, 
 $w_n$ is proportional to the probability of $n$ successive molecules, which are separated by less than $\tau_s$ time. 
No other molecules are present before and after this `\textit{molecular train}' within $\tau_S$. 
In case of random arrival of the molecules, the probability density of $\tau$ before the arrival of the next molecule is $\tau_m^{-1}\exp(\tau/\tau_m)$. 
For the probability of $\tau > \tau_s$ is $\exp(-\tau_s/\tau_m)$, which results to the weighting factor as
\begin{equation}
    \begin{split}
    w_n = \exp\Bigg(\!\!\!-\frac{\tau_s}{\tau_m}\!\!\Bigg)\Bigg[\prod_{j=1}^{n-1}\int^{\tau_s}_0  \!\!\!\exp\Bigg(-\frac{\tau_j}{\tau_m}\Bigg)\frac{\text d\tau_j}{\tau_m}\Bigg]\exp\Bigg(\!\!\!-\frac{\tau_s}{\tau_m}\!\!\Bigg) \\
    =\exp\Bigg(-\frac{2\tau_s}{\tau_m}\Bigg)\Bigg[1-\exp\Bigg(-\frac{\tau_s}{\tau_m}\Bigg)\Bigg]^{n-1}
    \end{split}
\end{equation}
%
\noindent Here, we consider a constant value of $\tau_s$, which is modified in the algorithm with a threshold loop. 
The initial condition of the numerical simulation was fed with spatial dependence of the intensity of the excitation laser $I(\bm r)$, collection efficiency $\eta(\bm r)$, and initial distribution of the molecule $p_0(\bm r)$. 
The intensity distribution of a diffraction-limited confocal laser beam is Gaussian. 
The nanochannels width and height are much smaller than the { {diameter (as well as full width half maximum)}} of the focus but the length is not (Figure \ref{fig:3}). 
For simplicity, we could neglect the Gaussian distribution in two axes for the quasi-1D nature of molecules' presence inside nanochannel but we haven't in the actual algorithm. We  have considered it as:
\begin{equation}
    I(x,y) = \frac{2W}{\pi r_\omega^2} \exp \Bigg[-2\frac{x^2 - y^2}{r_\omega^2} \Bigg]
\end{equation}
where $W$ denotes the power of the laser and $r_\omega$ is the waist radius. The spatial dependence of optical collection efficiency is given by (molecules are flowing along the $x$ axis):
\begin{multline}
    \eta(x,y) = \frac{\eta_0}{\pi(1-\cos\psi)} 
    \times \Bigg[\arcsin{\frac{\sin\theta}{\cos\psi}}\\
    - \cos\psi\arctan\Bigg(\frac{\cos\psi \sin\theta}{\sqrt{\sin^2\psi - \sin^2\theta}} \Bigg) \Bigg]_{\theta_{\min}}^{\theta_{\max}} 
\end{multline}
\noindent where $\eta_0$ is the maximum value of the collection efficiency, $\psi = \arcsin(\text{NA}/\underline{n})$, $\theta_{max} = \max{(-\arctan((\frac{d}{2}-x)/|y|), -\psi)}$ and $\theta_{min} = \max{(-\arctan((\frac{d}{2}-x)/|y|), -\psi)}$, NA is the value of the numerical aperture, $\underline{n}$ is the refractive index, and $d$ is the pinhole diameter in the object space. Here, $\theta_{\min} < \theta_{\max}$; if $\theta_{\min} > \theta_{\max}$, then $\eta(x,y)$ will be zero.

Employing a simple hydrodynamic model, we calculate the distribution of the molecules at a position $x=x_0$ outside the slice of the confocal volume with diffraction-limited focus.  
We have neglected the third dimension due to the confinement. 
Molecules are considered to be uniformly distributed over $x + c = y$; $c$ is a spatial constant. 
From this starting plane, molecules accelerate to the laminar flow velocity and undergo a confined diffusion. The starting point $x=x_0$ of the Monte-Carlo simulation is considered at $-3\omega$ where the light intensity is negligible. 
Then, the initial probability distribution $p_0$ is given by: 
\begin{equation}
    p_0(x,y) = \frac{x-x_0}{4\pi^2 y D \tau_0} \int_0^y \text{d}\bm{r} \int_0^{2\pi} \text{d}\phi \times \exp \Bigg[- \frac{(y - \bm{r} \sin\phi)^2}{4D\tau_0} \Bigg]
\end{equation}

\noindent$\tau_0$ is determined with respect to the pulsed frequency and the following equation:
\begin{equation}
    \frac{x_0 - x_{\text{inj}}}{v_0} = \tau_0 - k^{-1}[1- \exp(k\tau_0)]
    \vspace{-1mm}
\end{equation}
$k$ is an empirical flow acceleration constant. 

\section*{Single-molecule and Multi-molecule Bursts}

 {Earlier, we reported a detailed single-molecule nanofluidic flow and diffusion inside 30 nm to 100 nm all-silica nanofluidic channels\cite{ghosh2020single}}. 
We use a similar experimental configuration as shown in Figure \ref{fig:3}a (and Supplementary Figure 2) -- single molecules are flowing inside a nanochannel ({ {with cross-sectional diameter 50 nm -- in between the earlier experimental range))}} that is placed at the centre of two 640 nm  {laser} foci. 
{ {The green and violet foci represent vertical and horizontal polarisation, respectively from the optical axis. Time-tagged photon arrival to detectors from each focus let us identify the source of the photon as shown in Fig 1d.}} 
They are spatially separated but partially overlapping with $\approx$ 50 nm, { which is typically seen in a two-foci fluorescence correlation spectroscopy setups \cite{dertinger2007two, ghosh2020single}.}

The diffusion coefficients of  {1.127 nm} and  {11.27 nm} sized single molecules are 218 $\upmu$m$^2/$s and 21.8 $\upmu$m$^2/$s, respectively,  {and the size ratio between molecules and the chosen cross-sectional diameter of nanofluidic channel (molecule:nanochannel) are 0.022 and 0.22, respectively.} 
The time step used in the Monte-Carlo simulation was $\Delta\tau = 50~\upmu\text{s}$, $T$ was set to 3 ms  {(which are sufficiently smaller than experimental values to obtain detailed understanding)}, and the simulations sampled $10^4$ paths.
 {The chosen values are suitable for a regular eight core CPU to obtain results within several tens of minutes. We used an Intel Core i7-8650U CPU at 1.90 GHz × 8 with 16 GB RAM.}
 {A free-flowing or diffusing single molecule flowing through these two foci produces two bursts as schematically shown in Figure \ref{fig:3}a with green and violet time trace signals.
In Figure \ref{fig:3}(c), we show FEPI-MC simulated single-molecule bursts of  {1.127 nm}-sized molecules. }
The arrows in the time traces show some exemplary single-molecule bursts. 
The first arrow in Figure \ref{fig:3}(c)i represent single-molecule bursts with minimal interaction in both the foci. 
Bursts of same molecules from the two foci should produce same time duration. 
Molecules can slow down due to crawling effect. 
If there are multiple molecules present in the foci with close successions as shown in Figure \ref{fig:3}(b). 
Surface-adsorbed molecules (can be photobleached) are fixed sources of interaction with the flowing molecules. 
In Figure \ref{fig:3}(c)ii, we found a single-molecule burst with high intensity/photon numbers, which also shows a crawling effect as shown by the first arrow.  
Significantly long reappearing bursts within 250 ms is shown in Figure \ref{fig:3}(c)iii -- photo-inactive behaviour due to multi-molecule interactions. 
Another multi-molecule interaction for nearly 200 ms is shown in multi-molecule interactions. 
To distinguish between crawling or surface diffusion of single molecules \cite{walder2011single, wang2020non} on the nanochannel wall due to confined flow and multi-molecule interactions, we should study the electrodynamic interactions of single-molecule fluorescence (Supplementary Figure 4). 
 {The FEPI-MC simulated bursts resembles with the burst sizes and shapes of experimental time-trace  of free-flowing or diffusing single molecules inside a nanofluidic channel
For comparison, we show experimental single-molecule time-trace of Atto-488 carboxy molecules ($\approx$1.22 nm big) and 48 bp DNA tagged with Alexa 647 ($\approx$11.27 nm big) in Figure \ref{fig:3}(d). 
The characteristic size ratio of molecule:nanochannel is approximately 0.022 for Atto-488 in 50 nm cross-sectional channels and approximately 0.36 for 48bp DNAs flowing through a 30 nm nanochannel. 
Achieving perfect zero velocity in experiment is not straightforward as leakages and evaporation are almost unavoidable.
Further measurement details of the Figure \ref{fig:3}(d) can be found in \cite{ghosh2020single}}.

\begin{figure*}[t]
    \centering
    \includegraphics[width=0.8\textwidth]{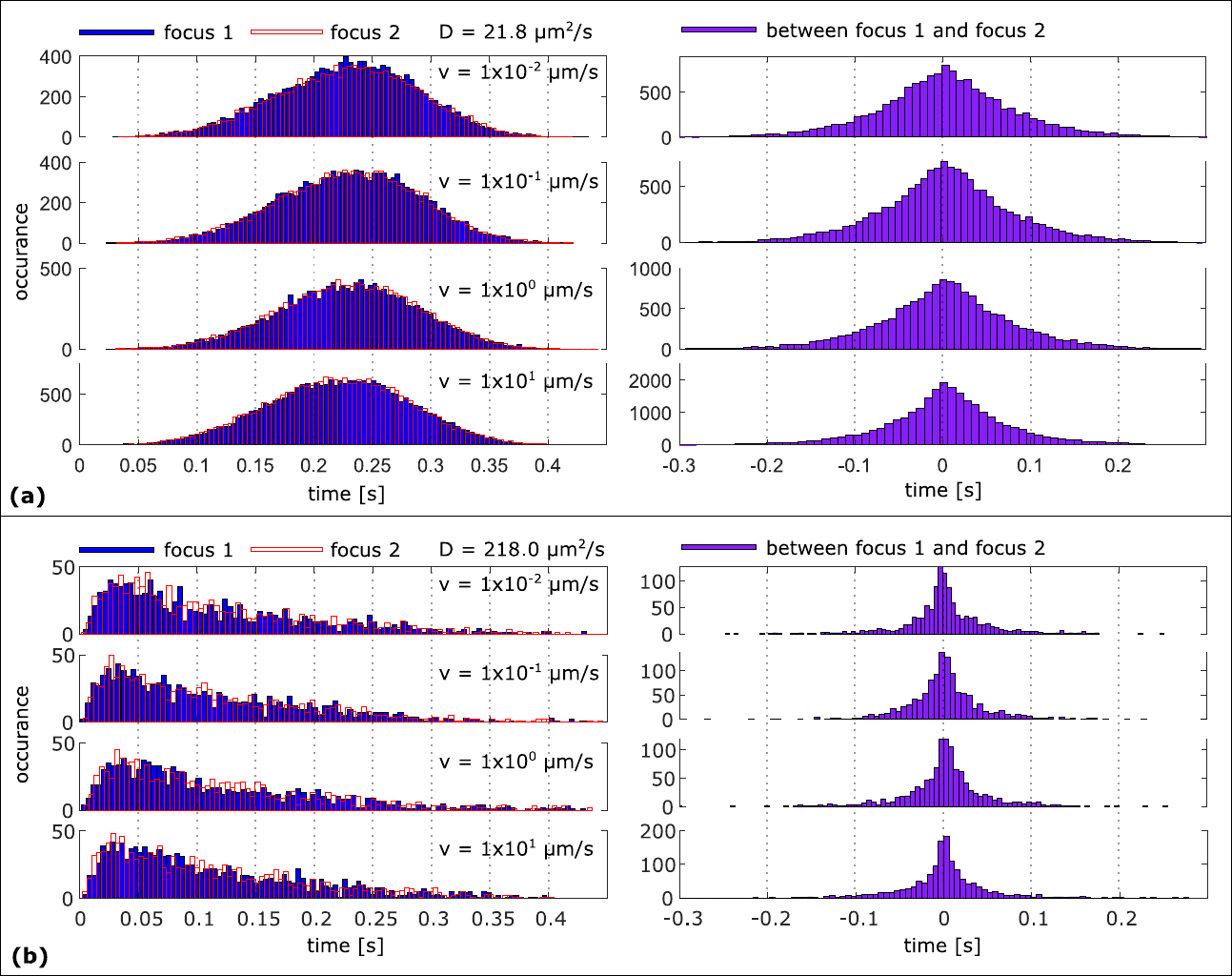}
    \vspace{-3mm}
    \caption{\textbf{Single-molecule BSD.} LHS: Histograms of (a)  {11.27 nm} and (b)  {1.127 nm} sized single molecules flowing through focus 1 and focus 2. 
    RHS: Histogram of time distribution between focus 1 and focus 2.}
    \label{fig:5}
    \vspace{-4mm}
\end{figure*}

\section*{Temporal distribution of bursts}
The FEPI-MC considers a $\delta$-function functional considering the complex interactions of single-molecule fluorescence experiments inside the nanofluidic channel as shown in equation 3. 
The histogram of the temporal response of single-molecule bursts will show the statistical nature of the complex interactions. 
Histograms of single-molecule burst size (defined with time) for  {11.27 nm}-sized single molecules from $10^4$ paths through focus 1 and 2 at
different velocities ranging from $0.01~\upmu$m/s to $10~\upmu$m/s is shown in In Figure \ref{fig:5}a. 
The histograms have broad time distributions of approximately 400 ms. 
At slow velocities, $0.01~\upmu$m/s and $0.1~\upmu$m/s, we find discreet peaks within the broad time distribution due to favourable interactions due to nanoconfined photophysical interactions of single molecules. 
Here, 210 ms to 230 ms single-molecule bursts are most frequently occurring, such as 230 ms for $0.01~\upmu$m/s has, 220 ms for $0.1~\upmu$m/s, 210 ms for $1~\upmu$m/s, and 200 ms to 250 ms for 10 $\upmu$m/s.
A 50 ms flat peak is observed at 10 $\upmu$m/s. 
These show the possibility of several kinds of interaction. 
Since we have two sources of signal for a single molecule, we compare the time distribution between focus 1 and focus 2 as well. 
The spread between two foci reduces from 400 ms to 200 ms with increasing velocities as shown in the right panel of Figure \ref{fig:5}a. 

For  {1.127 nm} size, the peak of the histogram shifts to a faster timescale of 25 -- 50 ms and the shape of the histogram is positively skewed as shown in Figure \ref{fig:5}b. 
The histogram for $0.01~\upmu$m/s shows high occurrences at fast timescale 25 ms, 60 ms, and 80 ms along with several low occurrences of long bursts at slow timescale up to 400 ms. 
The burst distributions at the faster timescale is due to single-molecule bursts, which have negligible wall interactions and passed through the foci with short intervals. 
The sub-25 ms timescale bursts are non-interacting single-molecule crossings, and the 50 to 100 ms timescale bursts are responsible for molecular shot noise i.e.~molecular interactions among multiple single molecules. 
The slower occurrences are due to single-molecule interactions with the walls --- crawling events of single-molecule or multi-molecules. 
At slow velocities, the single-molecule crossings, molecular shot-noise, and crawling events can be well identified, which get smeared away at fast velocities as shown in Figure \ref{fig:5}b.
The time distribution between focus 1 and focus 2 is narrower at 10 $\upmu$m/s compare to slow flow velocities as shown in the right panel of Figure \ref{fig:5}b. 
The time distributions also show fine temporal following the analysis timescale of different single-molecule bursts characteristics. 

\begin{figure*}[t]
    \centering
    \includegraphics[width=\textwidth]{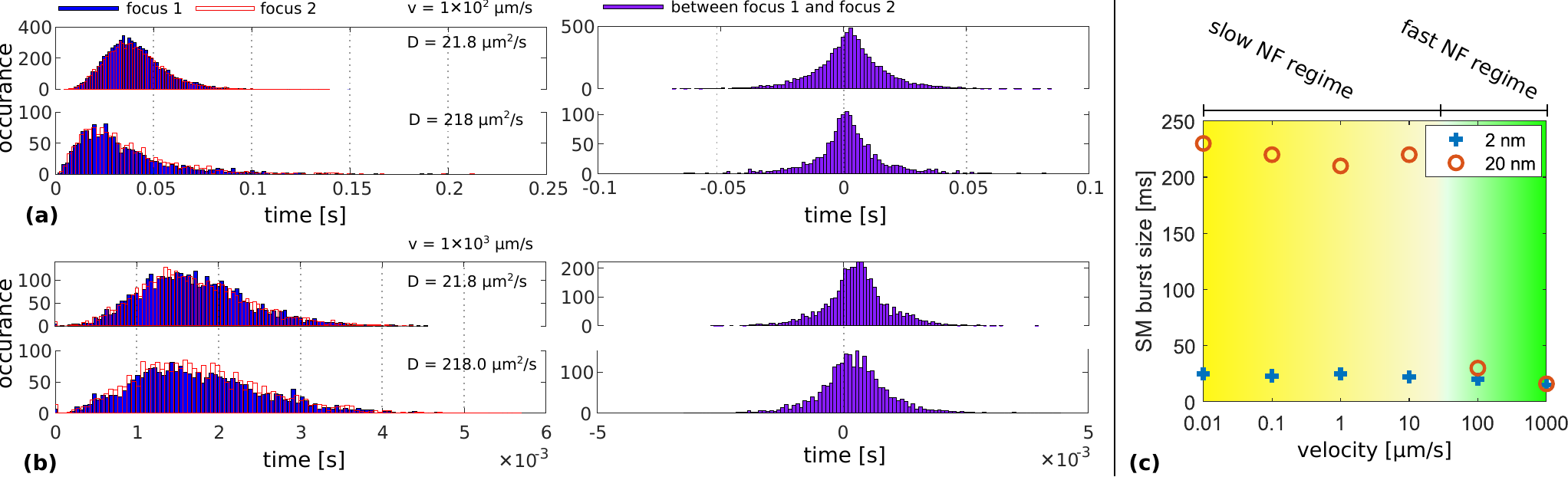}
    \vspace{-3mm}
    \caption{\textbf{BSD of fast nanofluidics regime}. \textbf{(a)} At flow velocity 100 $\upmu$m/s,  {11.27 nm} (D = 21.8 $\upmu$m$^2$/s) and  {1.127 nm} (D = 218 $\upmu$m$^2$/s). \textbf{(b)} At 1000 $\upmu$m/s flow velocity,  {11.27 nm} molecule and  {1.127 nm}.
    \textbf{(c)} Single-molecule nanofluidic regimes  {-- the relationship between single-molecule burst size and flow velocity showing the effect of nanofluidic regimes on single-molecule bursts;  {11.27 nm} single molecules are more visible compare to  {1.127 nm} single molecules in slow NF regime whereas at fast NF regime both behave similarly.}} 
    \label{fig:6}
    \vspace{-5mm}
\end{figure*}

We classify the flow velocities ranging from $0.01~\upmu$m/s to $10~\upmu$m/s as `slow nanofluidic regime' due to the dominating characteristic features of the single-molecule diffusion behaviour and their strong interactions for the nanofluidic confinement. 
Irrespective of the order of magnitude differences in flow velocities, we can identify reproducible slow nanofluidic regime characteristics. 
 {The  {11.27 nm} molecules show single-molecule bursts at around 220 ms, and at sub-25 ms or one order of magnitude faster timescale for  {1.127 nm} molecules' single molecule bursts. 
This suggests us that the single-molecule BSDs are potentially following a similar trend of dependency of diffusion coefficient to hydrodynamic radius of the molecule.}
The broad spread of occurrence in  {11.27 nm} single molecules at this nanofluidic regime show strong interactions/crawling effect and molecular shot noise. 

`Fast nanofluidic regime' is from $100~\upmu$m/s velocity as shown in Figure \ref{fig:6}. 
Figure \ref{fig:6}a shows histograms of  {11.27 nm} and  {1.127 nm} single molecules at $100~\upmu$m/s with the highest occurrence at 33 ms and 20 ms, respectively. 
At this flow regime, the aforementioned interactions of single molecules are not strong, specifically for the  {11.27 nm} molecules. 
The molecular shot noise and crawling events are limited. 
However, for the  {1.127 nm} molecules, the molecular shot noise scales down to a faster time scale as we can see a distribution of this peak at 38 ms along with a small contribution due to crawling events.  
Sub-20 ms bursts for  {1.127 nm} molecules are consistent at a fast nanofluidic regime. 
The time distributions between two foci also suggest the same in the right panel.  
The trend continues to $1000~\upmu$m/s in Figure \ref{fig:6}b. 
At this flow velocity, it is hard to distinguish between the histograms of  {11.27 nm} and  {1.127 nm} single molecules, except for the fine features. 
At the fast nanofluidic regime, the driving force for flow is large enough to overcome the crawling events. 
For an overall understanding, Figure \ref{fig:6}c shows single-molecule burst sizes of the two identified nanofluidic regimes with respect to the flow velocities.

\section*{Conclusion}
The Feynman-Enderlein path integral is a powerful method to model the complex single-molecule nanofluidics and show two distinct single-molecule nanofluidic regimes. 
We have shown how to resolve the complexity of single-molecule nanofluidics by integrating electrodynamics. 
Electrodynamic interaction of the electronic structure of a single molecule significantly influences the decay time of fluorescence.  
 {Fluorescence lifetime is highly sensitive to the interface. 
Thus, single-molecule nanofluidic experiments inside non-uniform interfaces, such as silicon-silica or silica-PDMS \cite{Ghosh2014-1,Hao2017,kuhn2006modification,kuhn2008modification,liaw2009purcell} most of the time do not reproduce single molecule fluorescence signals.}
In future, the method will be integrated with the electrostatic effects to deal with the Debye-length related issues. 
The FEPI-MC method has not been used widely in single-molecule experiments. 
This method opens up several avenues in biophysics as well as quantum hydrodynamics. 
Single fluorophores to misfolded proteins, nanobodies, and quantum dots in solid-state nanochannels/tunnelling nanotubes or within lipid bilayers are all relevant for this work. 
Most importantly, the exact formulation of the molecules is significantly close to the physics of single-molecule fluorescence and is open to include any new condition. 
The potential to exploit this direct method to predict as well as model/readout of single-molecule or particle dynamics (where many-body interactions is strong) opens nanofluidics, which cannot be used at its fullest due to the unavailability of this kind of straightforward mathematical approach. 
A machine learning-based platform using FEPI is a future avenue to make this method available to the unrestricted communities of chemical biology, lab-on-a-chip, optofluidic, and fluid dynamics of many-body interactions.   

\section*{Conflicts of interest}
There are no conflicts to declare.

\section*{Acknowledgement}
The research was funded by the German Research Foundation/Deutsche Forschungsgemeinschaft (DFG) -- Project number 405479535 (PI Siddharth Ghosh). 
The author is grateful to Professor Joerg Enderlein for many important discussions and for laying the foundation of this work. 
Computational resources used here are provided by Professor Tuomas Knowles at the University of Cambridge.

\bibliography{Ref.bib} 

\begin{thebibliography}{50}%
\makeatletter
\providecommand \@ifxundefined [1]{%
 \@ifx{#1\undefined}
}%
\providecommand \@ifnum [1]{%
 \ifnum #1\expandafter \@firstoftwo
 \else \expandafter \@secondoftwo
 \fi
}%
\providecommand \@ifx [1]{%
 \ifx #1\expandafter \@firstoftwo
 \else \expandafter \@secondoftwo
 \fi
}%
\providecommand \natexlab [1]{#1}%
\providecommand \enquote  [1]{``#1''}%
\providecommand \bibnamefont  [1]{#1}%
\providecommand \bibfnamefont [1]{#1}%
\providecommand \citenamefont [1]{#1}%
\providecommand \href@noop [0]{\@secondoftwo}%
\providecommand \href [0]{\begingroup \@sanitize@url \@href}%
\providecommand \@href[1]{\@@startlink{#1}\@@href}%
\providecommand \@@href[1]{\endgroup#1\@@endlink}%
\providecommand \@sanitize@url [0]{\catcode `\\12\catcode `\$12\catcode
  `\&12\catcode `\#12\catcode `\^12\catcode `\_12\catcode `\%12\relax}%
\providecommand \@@startlink[1]{}%
\providecommand \@@endlink[0]{}%
\providecommand \url  [0]{\begingroup\@sanitize@url \@url }%
\providecommand \@url [1]{\endgroup\@href {#1}{\urlprefix }}%
\providecommand \urlprefix  [0]{URL }%
\providecommand \Eprint [0]{\href }%
\providecommand \doibase [0]{http://dx.doi.org/}%
\providecommand \selectlanguage [0]{\@gobble}%
\providecommand \bibinfo  [0]{\@secondoftwo}%
\providecommand \bibfield  [0]{\@secondoftwo}%
\providecommand \translation [1]{[#1]}%
\providecommand \BibitemOpen [0]{}%
\providecommand \bibitemStop [0]{}%
\providecommand \bibitemNoStop [0]{.\EOS\space}%
\providecommand \EOS [0]{\spacefactor3000\relax}%
\providecommand \BibitemShut  [1]{\csname bibitem#1\endcsname}%
\let\auto@bib@innerbib\@empty
\bibitem [{\citenamefont {Schroedinger}(2012)}]{schrodinger2012life}%
  \BibitemOpen
  \bibfield  {author} {\bibinfo {author} {\bibfnamefont {E.}~\bibnamefont
  {Schroedinger}},\ }\href@noop {} {\emph {\bibinfo {title} {What is life?:
  With mind and matter and autobiographical sketches}}}\ (\bibinfo  {publisher}
  {Cambridge University Press},\ \bibinfo {year} {2012})\BibitemShut {NoStop}%
\bibitem [{\citenamefont {Loose}\ \emph {et~al.}(2008)\citenamefont {Loose},
  \citenamefont {Fischer-Friedrich}, \citenamefont {Ries}, \citenamefont
  {Kruse},\ and\ \citenamefont {Schwille}}]{loose2008spatial}%
  \BibitemOpen
  \bibfield  {author} {\bibinfo {author} {\bibfnamefont {M.}~\bibnamefont
  {Loose}}, \bibinfo {author} {\bibfnamefont {E.}~\bibnamefont
  {Fischer-Friedrich}}, \bibinfo {author} {\bibfnamefont {J.}~\bibnamefont
  {Ries}}, \bibinfo {author} {\bibfnamefont {K.}~\bibnamefont {Kruse}}, \ and\
  \bibinfo {author} {\bibfnamefont {P.}~\bibnamefont {Schwille}},\ }\href@noop
  {} {\bibfield  {journal} {\bibinfo  {journal} {Science}\ }\textbf {\bibinfo
  {volume} {320}},\ \bibinfo {pages} {789} (\bibinfo {year}
  {2008})}\BibitemShut {NoStop}%
\bibitem [{\citenamefont {Litschel}\ \emph {et~al.}(2018)\citenamefont
  {Litschel}, \citenamefont {Ganzinger}, \citenamefont {Movinkel},
  \citenamefont {Heymann}, \citenamefont {Robinson}, \citenamefont
  {Mutschler},\ and\ \citenamefont {Schwille}}]{litschel2018freeze}%
  \BibitemOpen
  \bibfield  {author} {\bibinfo {author} {\bibfnamefont {T.}~\bibnamefont
  {Litschel}}, \bibinfo {author} {\bibfnamefont {K.~A.}\ \bibnamefont
  {Ganzinger}}, \bibinfo {author} {\bibfnamefont {T.}~\bibnamefont {Movinkel}},
  \bibinfo {author} {\bibfnamefont {M.}~\bibnamefont {Heymann}}, \bibinfo
  {author} {\bibfnamefont {T.}~\bibnamefont {Robinson}}, \bibinfo {author}
  {\bibfnamefont {H.}~\bibnamefont {Mutschler}}, \ and\ \bibinfo {author}
  {\bibfnamefont {P.}~\bibnamefont {Schwille}},\ }\href@noop {} {\bibfield
  {journal} {\bibinfo  {journal} {New Journal of Physics}\ }\textbf {\bibinfo
  {volume} {20}},\ \bibinfo {pages} {055008} (\bibinfo {year}
  {2018})}\BibitemShut {NoStop}%
\bibitem [{\citenamefont {Golestanian}(2015)}]{golestanian2015enhanced}%
  \BibitemOpen
  \bibfield  {author} {\bibinfo {author} {\bibfnamefont {R.}~\bibnamefont
  {Golestanian}},\ }\href@noop {} {\bibfield  {journal} {\bibinfo  {journal}
  {Physical review letters}\ }\textbf {\bibinfo {volume} {115}},\ \bibinfo
  {pages} {108102} (\bibinfo {year} {2015})}\BibitemShut {NoStop}%
\bibitem [{\citenamefont {Illien}\ \emph {et~al.}(2017)\citenamefont {Illien},
  \citenamefont {Zhao}, \citenamefont {Dey}, \citenamefont {Butler},
  \citenamefont {Sen},\ and\ \citenamefont
  {Golestanian}}]{illien2017exothermicity}%
  \BibitemOpen
  \bibfield  {author} {\bibinfo {author} {\bibfnamefont {P.}~\bibnamefont
  {Illien}}, \bibinfo {author} {\bibfnamefont {X.}~\bibnamefont {Zhao}},
  \bibinfo {author} {\bibfnamefont {K.~K.}\ \bibnamefont {Dey}}, \bibinfo
  {author} {\bibfnamefont {P.~J.}\ \bibnamefont {Butler}}, \bibinfo {author}
  {\bibfnamefont {A.}~\bibnamefont {Sen}}, \ and\ \bibinfo {author}
  {\bibfnamefont {R.}~\bibnamefont {Golestanian}},\ }\href@noop {} {\bibfield
  {journal} {\bibinfo  {journal} {Nano letters}\ }\textbf {\bibinfo {volume}
  {17}},\ \bibinfo {pages} {4415} (\bibinfo {year} {2017})}\BibitemShut
  {NoStop}%
\bibitem [{\citenamefont {Ghosh}\ \emph
  {et~al.}(2016{\natexlab{a}})\citenamefont {Ghosh}, \citenamefont {Ghosh},
  \citenamefont {Seibt}, \citenamefont {Rao}, \citenamefont {Peretzki},\ and\
  \citenamefont {Rao}}]{ghosh2016ferroelectric}%
  \BibitemOpen
  \bibfield  {author} {\bibinfo {author} {\bibfnamefont {M.}~\bibnamefont
  {Ghosh}}, \bibinfo {author} {\bibfnamefont {S.}~\bibnamefont {Ghosh}},
  \bibinfo {author} {\bibfnamefont {M.}~\bibnamefont {Seibt}}, \bibinfo
  {author} {\bibfnamefont {K.~Y.}\ \bibnamefont {Rao}}, \bibinfo {author}
  {\bibfnamefont {P.}~\bibnamefont {Peretzki}}, \ and\ \bibinfo {author}
  {\bibfnamefont {G.~M.}\ \bibnamefont {Rao}},\ }\href@noop {} {\bibfield
  {journal} {\bibinfo  {journal} {CrystEngComm}\ }\textbf {\bibinfo {volume}
  {18}},\ \bibinfo {pages} {622} (\bibinfo {year}
  {2016}{\natexlab{a}})}\BibitemShut {NoStop}%
\bibitem [{\citenamefont {Ghosh}\ \emph
  {et~al.}(2016{\natexlab{b}})\citenamefont {Ghosh}, \citenamefont {Ghosh},
  \citenamefont {Attariani}, \citenamefont {Momeni}, \citenamefont {Seibt},\
  and\ \citenamefont {Mohan~Rao}}]{ghosh2016atomic}%
  \BibitemOpen
  \bibfield  {author} {\bibinfo {author} {\bibfnamefont {M.}~\bibnamefont
  {Ghosh}}, \bibinfo {author} {\bibfnamefont {S.}~\bibnamefont {Ghosh}},
  \bibinfo {author} {\bibfnamefont {H.}~\bibnamefont {Attariani}}, \bibinfo
  {author} {\bibfnamefont {K.}~\bibnamefont {Momeni}}, \bibinfo {author}
  {\bibfnamefont {M.}~\bibnamefont {Seibt}}, \ and\ \bibinfo {author}
  {\bibfnamefont {G.}~\bibnamefont {Mohan~Rao}},\ }\href@noop {} {\bibfield
  {journal} {\bibinfo  {journal} {Nano letters}\ }\textbf {\bibinfo {volume}
  {16}},\ \bibinfo {pages} {5969} (\bibinfo {year}
  {2016}{\natexlab{b}})}\BibitemShut {NoStop}%
\bibitem [{\citenamefont
  {Enderlein}(2000{\natexlab{a}})}]{enderlein2000positional}%
  \BibitemOpen
  \bibfield  {author} {\bibinfo {author} {\bibfnamefont {J.}~\bibnamefont
  {Enderlein}},\ }\href@noop {} {\bibfield  {journal} {\bibinfo  {journal}
  {Single Molecules}\ }\textbf {\bibinfo {volume} {1}},\ \bibinfo {pages} {225}
  (\bibinfo {year} {2000}{\natexlab{a}})}\BibitemShut {NoStop}%
\bibitem [{\citenamefont {Cohen}(2007)}]{Cohen2006}%
  \BibitemOpen
  \bibfield  {author} {\bibinfo {author} {\bibfnamefont {A.~E.}\ \bibnamefont
  {Cohen}},\ }\emph {\bibinfo {title} {Trapping and manipulating single
  molecules in solution}},\ \href@noop {} {Ph.D. thesis},\ \bibinfo  {school}
  {Stanford University} (\bibinfo {year} {2007})\BibitemShut {NoStop}%
\bibitem [{\citenamefont {Cohen}\ and\ \citenamefont
  {Moerner}(2006)}]{CohenPNAS2006}%
  \BibitemOpen
  \bibfield  {author} {\bibinfo {author} {\bibfnamefont {A.~E.}\ \bibnamefont
  {Cohen}}\ and\ \bibinfo {author} {\bibfnamefont {W.}~\bibnamefont
  {Moerner}},\ }\href@noop {} {\bibfield  {journal} {\bibinfo  {journal}
  {Proceedings of the National Academy of Sciences}\ }\textbf {\bibinfo
  {volume} {103}},\ \bibinfo {pages} {4362} (\bibinfo {year}
  {2006})}\BibitemShut {NoStop}%
\bibitem [{\citenamefont {Ghosh}\ \emph {et~al.}(2020)\citenamefont {Ghosh},
  \citenamefont {Karedla},\ and\ \citenamefont {Gregor}}]{ghosh2020single}%
  \BibitemOpen
  \bibfield  {author} {\bibinfo {author} {\bibfnamefont {S.}~\bibnamefont
  {Ghosh}}, \bibinfo {author} {\bibfnamefont {N.}~\bibnamefont {Karedla}}, \
  and\ \bibinfo {author} {\bibfnamefont {I.}~\bibnamefont {Gregor}},\
  }\href@noop {} {\bibfield  {journal} {\bibinfo  {journal} {Lab on a Chip}\
  }\textbf {\bibinfo {volume} {20}},\ \bibinfo {pages} {3249} (\bibinfo {year}
  {2020})}\BibitemShut {NoStop}%
\bibitem [{\citenamefont {Ranzinger}\ \emph {et~al.}(2011)\citenamefont
  {Ranzinger}, \citenamefont {Rustom}, \citenamefont {Abel}, \citenamefont
  {Leyh}, \citenamefont {Kihm}, \citenamefont {Witkowski}, \citenamefont
  {Scheurich}, \citenamefont {Zeier},\ and\ \citenamefont
  {Schwenger}}]{ranzinger2011nanotube}%
  \BibitemOpen
  \bibfield  {author} {\bibinfo {author} {\bibfnamefont {J.}~\bibnamefont
  {Ranzinger}}, \bibinfo {author} {\bibfnamefont {A.}~\bibnamefont {Rustom}},
  \bibinfo {author} {\bibfnamefont {M.}~\bibnamefont {Abel}}, \bibinfo {author}
  {\bibfnamefont {J.}~\bibnamefont {Leyh}}, \bibinfo {author} {\bibfnamefont
  {L.}~\bibnamefont {Kihm}}, \bibinfo {author} {\bibfnamefont {M.}~\bibnamefont
  {Witkowski}}, \bibinfo {author} {\bibfnamefont {P.}~\bibnamefont
  {Scheurich}}, \bibinfo {author} {\bibfnamefont {M.}~\bibnamefont {Zeier}}, \
  and\ \bibinfo {author} {\bibfnamefont {V.}~\bibnamefont {Schwenger}},\
  }\href@noop {} {\bibfield  {journal} {\bibinfo  {journal} {PLoS One}\
  }\textbf {\bibinfo {volume} {6}},\ \bibinfo {pages} {e29537} (\bibinfo {year}
  {2011})}\BibitemShut {NoStop}%
\bibitem [{\citenamefont {Jeffet}\ \emph {et~al.}(2016)\citenamefont {Jeffet},
  \citenamefont {Kobo}, \citenamefont {Su}, \citenamefont {Grunwald},
  \citenamefont {Green}, \citenamefont {Nilsson}, \citenamefont {Eisenberg},
  \citenamefont {Ambjornsson}, \citenamefont {Westerlund}, \citenamefont
  {Weinhold} \emph {et~al.}}]{jeffet2016super}%
  \BibitemOpen
  \bibfield  {author} {\bibinfo {author} {\bibfnamefont {J.}~\bibnamefont
  {Jeffet}}, \bibinfo {author} {\bibfnamefont {A.}~\bibnamefont {Kobo}},
  \bibinfo {author} {\bibfnamefont {T.}~\bibnamefont {Su}}, \bibinfo {author}
  {\bibfnamefont {A.}~\bibnamefont {Grunwald}}, \bibinfo {author}
  {\bibfnamefont {O.}~\bibnamefont {Green}}, \bibinfo {author} {\bibfnamefont
  {A.~N.}\ \bibnamefont {Nilsson}}, \bibinfo {author} {\bibfnamefont
  {E.}~\bibnamefont {Eisenberg}}, \bibinfo {author} {\bibfnamefont
  {T.}~\bibnamefont {Ambjornsson}}, \bibinfo {author} {\bibfnamefont
  {F.}~\bibnamefont {Westerlund}}, \bibinfo {author} {\bibfnamefont
  {E.}~\bibnamefont {Weinhold}},  \emph {et~al.},\ }\href@noop {} {\bibfield
  {journal} {\bibinfo  {journal} {ACS nano}\ }\textbf {\bibinfo {volume}
  {10}},\ \bibinfo {pages} {9823} (\bibinfo {year} {2016})}\BibitemShut
  {NoStop}%
\bibitem [{\citenamefont {Ristanovic}\ \emph {et~al.}(2018)\citenamefont
  {Ristanovic}, \citenamefont {Chowdhury}, \citenamefont {Brogaard},
  \citenamefont {Houben}, \citenamefont {Baldus}, \citenamefont {Hofkens},
  \citenamefont {Roeffaers},\ and\ \citenamefont
  {Weckhuysen}}]{zoran2018reversible}%
  \BibitemOpen
  \bibfield  {author} {\bibinfo {author} {\bibfnamefont {Z.}~\bibnamefont
  {Ristanovic}}, \bibinfo {author} {\bibfnamefont {A.~D.}\ \bibnamefont
  {Chowdhury}}, \bibinfo {author} {\bibfnamefont {R.~Y.}\ \bibnamefont
  {Brogaard}}, \bibinfo {author} {\bibfnamefont {K.}~\bibnamefont {Houben}},
  \bibinfo {author} {\bibfnamefont {M.}~\bibnamefont {Baldus}}, \bibinfo
  {author} {\bibfnamefont {J.}~\bibnamefont {Hofkens}}, \bibinfo {author}
  {\bibfnamefont {M.~B.}\ \bibnamefont {Roeffaers}}, \ and\ \bibinfo {author}
  {\bibfnamefont {B.~M.}\ \bibnamefont {Weckhuysen}},\ }\href@noop {}
  {\bibfield  {journal} {\bibinfo  {journal} {Journal of the American Chemical
  Society}\ }\textbf {\bibinfo {volume} {140}},\ \bibinfo {pages} {14195}
  (\bibinfo {year} {2018})}\BibitemShut {NoStop}%
\bibitem [{\citenamefont {Holt}\ \emph {et~al.}(2006)\citenamefont {Holt},
  \citenamefont {Park}, \citenamefont {Wang}, \citenamefont {Stadermann},
  \citenamefont {Artyukhin}, \citenamefont {Grigoropoulos}, \citenamefont
  {Noy},\ and\ \citenamefont {Bakajin}}]{Noy2006}%
  \BibitemOpen
  \bibfield  {author} {\bibinfo {author} {\bibfnamefont {J.~K.}\ \bibnamefont
  {Holt}}, \bibinfo {author} {\bibfnamefont {H.~G.}\ \bibnamefont {Park}},
  \bibinfo {author} {\bibfnamefont {Y.}~\bibnamefont {Wang}}, \bibinfo {author}
  {\bibfnamefont {M.}~\bibnamefont {Stadermann}}, \bibinfo {author}
  {\bibfnamefont {A.~B.}\ \bibnamefont {Artyukhin}}, \bibinfo {author}
  {\bibfnamefont {C.~P.}\ \bibnamefont {Grigoropoulos}}, \bibinfo {author}
  {\bibfnamefont {A.}~\bibnamefont {Noy}}, \ and\ \bibinfo {author}
  {\bibfnamefont {O.}~\bibnamefont {Bakajin}},\ }\href@noop {} {\bibfield
  {journal} {\bibinfo  {journal} {Science}\ }\textbf {\bibinfo {volume}
  {312}},\ \bibinfo {pages} {1034} (\bibinfo {year} {2006})}\BibitemShut
  {NoStop}%
\bibitem [{\citenamefont {Tunuguntla}\ \emph {et~al.}(2017)\citenamefont
  {Tunuguntla}, \citenamefont {Henley}, \citenamefont {Yao}, \citenamefont
  {Pham}, \citenamefont {Wanunu},\ and\ \citenamefont {Noy}}]{Noy2017}%
  \BibitemOpen
  \bibfield  {author} {\bibinfo {author} {\bibfnamefont {R.~H.}\ \bibnamefont
  {Tunuguntla}}, \bibinfo {author} {\bibfnamefont {R.~Y.}\ \bibnamefont
  {Henley}}, \bibinfo {author} {\bibfnamefont {Y.-C.}\ \bibnamefont {Yao}},
  \bibinfo {author} {\bibfnamefont {T.~A.}\ \bibnamefont {Pham}}, \bibinfo
  {author} {\bibfnamefont {M.}~\bibnamefont {Wanunu}}, \ and\ \bibinfo {author}
  {\bibfnamefont {A.}~\bibnamefont {Noy}},\ }\href@noop {} {\bibfield
  {journal} {\bibinfo  {journal} {Science}\ }\textbf {\bibinfo {volume}
  {357}},\ \bibinfo {pages} {792} (\bibinfo {year} {2017})}\BibitemShut
  {NoStop}%
\bibitem [{\citenamefont {Lu}\ \emph {et~al.}(1998)\citenamefont {Lu},
  \citenamefont {Xun},\ and\ \citenamefont {Xie}}]{lu1998single}%
  \BibitemOpen
  \bibfield  {author} {\bibinfo {author} {\bibfnamefont {H.~P.}\ \bibnamefont
  {Lu}}, \bibinfo {author} {\bibfnamefont {L.}~\bibnamefont {Xun}}, \ and\
  \bibinfo {author} {\bibfnamefont {X.~S.}\ \bibnamefont {Xie}},\ }\href@noop
  {} {\bibfield  {journal} {\bibinfo  {journal} {Science}\ }\textbf {\bibinfo
  {volume} {282}},\ \bibinfo {pages} {1877} (\bibinfo {year}
  {1998})}\BibitemShut {NoStop}%
\bibitem [{\citenamefont {Zijlstra}\ \emph {et~al.}(2012)\citenamefont
  {Zijlstra}, \citenamefont {Paulo},\ and\ \citenamefont
  {Orrit}}]{zijlstra2012optical}%
  \BibitemOpen
  \bibfield  {author} {\bibinfo {author} {\bibfnamefont {P.}~\bibnamefont
  {Zijlstra}}, \bibinfo {author} {\bibfnamefont {P.~M.}\ \bibnamefont {Paulo}},
  \ and\ \bibinfo {author} {\bibfnamefont {M.}~\bibnamefont {Orrit}},\
  }\href@noop {} {\bibfield  {journal} {\bibinfo  {journal} {Nature
  nanotechnology}\ }\textbf {\bibinfo {volume} {7}},\ \bibinfo {pages} {379}
  (\bibinfo {year} {2012})}\BibitemShut {NoStop}%
\bibitem [{\citenamefont {Baaske}\ \emph {et~al.}(2014)\citenamefont {Baaske},
  \citenamefont {Foreman},\ and\ \citenamefont {Vollmer}}]{baaske2014single}%
  \BibitemOpen
  \bibfield  {author} {\bibinfo {author} {\bibfnamefont {M.~D.}\ \bibnamefont
  {Baaske}}, \bibinfo {author} {\bibfnamefont {M.~R.}\ \bibnamefont {Foreman}},
  \ and\ \bibinfo {author} {\bibfnamefont {F.}~\bibnamefont {Vollmer}},\
  }\href@noop {} {\bibfield  {journal} {\bibinfo  {journal} {Nature
  nanotechnology}\ }\textbf {\bibinfo {volume} {9}},\ \bibinfo {pages} {933}
  (\bibinfo {year} {2014})}\BibitemShut {NoStop}%
\bibitem [{\citenamefont {Lesoine}\ \emph {et~al.}(2012)\citenamefont
  {Lesoine}, \citenamefont {Venkataraman}, \citenamefont {Maloney},
  \citenamefont {Dumont},\ and\ \citenamefont
  {Novotny}}]{lesoine2012nanochannel}%
  \BibitemOpen
  \bibfield  {author} {\bibinfo {author} {\bibfnamefont {J.~F.}\ \bibnamefont
  {Lesoine}}, \bibinfo {author} {\bibfnamefont {P.~A.}\ \bibnamefont
  {Venkataraman}}, \bibinfo {author} {\bibfnamefont {P.~C.}\ \bibnamefont
  {Maloney}}, \bibinfo {author} {\bibfnamefont {M.~E.}\ \bibnamefont {Dumont}},
  \ and\ \bibinfo {author} {\bibfnamefont {L.}~\bibnamefont {Novotny}},\
  }\href@noop {} {\bibfield  {journal} {\bibinfo  {journal} {Nano letters}\
  }\textbf {\bibinfo {volume} {12}},\ \bibinfo {pages} {3273} (\bibinfo {year}
  {2012})}\BibitemShut {NoStop}%
\bibitem [{\citenamefont {Purcell}(1995)}]{purcell1995spontaneous}%
  \BibitemOpen
  \bibfield  {author} {\bibinfo {author} {\bibfnamefont {E.~M.}\ \bibnamefont
  {Purcell}},\ }in\ \href@noop {} {\emph {\bibinfo {booktitle} {Confined
  Electrons and Photons}}}\ (\bibinfo  {publisher} {Springer},\ \bibinfo {year}
  {1995})\ pp.\ \bibinfo {pages} {839--839}\BibitemShut {NoStop}%
\bibitem [{\citenamefont
  {Enderlein}(2000{\natexlab{b}})}]{enderlein2000theoretical}%
  \BibitemOpen
  \bibfield  {author} {\bibinfo {author} {\bibfnamefont {J.}~\bibnamefont
  {Enderlein}},\ }\href@noop {} {\bibfield  {journal} {\bibinfo  {journal}
  {Biophysical Journal}\ }\textbf {\bibinfo {volume} {78}},\ \bibinfo {pages}
  {2151} (\bibinfo {year} {2000}{\natexlab{b}})}\BibitemShut {NoStop}%
\bibitem [{\citenamefont {Karedla}\ \emph {et~al.}(2015)\citenamefont
  {Karedla}, \citenamefont {Stein}, \citenamefont {H{\"a}hnel}, \citenamefont
  {Gregor}, \citenamefont {Chizhik},\ and\ \citenamefont
  {Enderlein}}]{karedla2015simultaneous}%
  \BibitemOpen
  \bibfield  {author} {\bibinfo {author} {\bibfnamefont {N.}~\bibnamefont
  {Karedla}}, \bibinfo {author} {\bibfnamefont {S.~C.}\ \bibnamefont {Stein}},
  \bibinfo {author} {\bibfnamefont {D.}~\bibnamefont {H{\"a}hnel}}, \bibinfo
  {author} {\bibfnamefont {I.}~\bibnamefont {Gregor}}, \bibinfo {author}
  {\bibfnamefont {A.}~\bibnamefont {Chizhik}}, \ and\ \bibinfo {author}
  {\bibfnamefont {J.}~\bibnamefont {Enderlein}},\ }\href@noop {} {\bibfield
  {journal} {\bibinfo  {journal} {Physical review letters}\ }\textbf {\bibinfo
  {volume} {115}},\ \bibinfo {pages} {173002} (\bibinfo {year}
  {2015})}\BibitemShut {NoStop}%
\bibitem [{\citenamefont {Westerlund}\ \emph {et~al.}(2010)\citenamefont
  {Westerlund}, \citenamefont {Persson}, \citenamefont {Kristensen},\ and\
  \citenamefont {Tegenfeldt}}]{Tegenfeldt2010}%
  \BibitemOpen
  \bibfield  {author} {\bibinfo {author} {\bibfnamefont {F.}~\bibnamefont
  {Westerlund}}, \bibinfo {author} {\bibfnamefont {F.}~\bibnamefont {Persson}},
  \bibinfo {author} {\bibfnamefont {A.}~\bibnamefont {Kristensen}}, \ and\
  \bibinfo {author} {\bibfnamefont {J.~O.}\ \bibnamefont {Tegenfeldt}},\
  }\href@noop {} {\bibfield  {journal} {\bibinfo  {journal} {Lab on a Chip}\
  }\textbf {\bibinfo {volume} {10}},\ \bibinfo {pages} {2049} (\bibinfo {year}
  {2010})}\BibitemShut {NoStop}%
\bibitem [{\citenamefont {Eggeling}\ \emph {et~al.}(2001)\citenamefont
  {Eggeling}, \citenamefont {Berger}, \citenamefont {Brand}, \citenamefont
  {Fries}, \citenamefont {Schaffer}, \citenamefont {Volkmer},\ and\
  \citenamefont {Seidel}}]{eggeling2001data}%
  \BibitemOpen
  \bibfield  {author} {\bibinfo {author} {\bibfnamefont {C.}~\bibnamefont
  {Eggeling}}, \bibinfo {author} {\bibfnamefont {S.}~\bibnamefont {Berger}},
  \bibinfo {author} {\bibfnamefont {L.}~\bibnamefont {Brand}}, \bibinfo
  {author} {\bibfnamefont {J.}~\bibnamefont {Fries}}, \bibinfo {author}
  {\bibfnamefont {J.}~\bibnamefont {Schaffer}}, \bibinfo {author}
  {\bibfnamefont {A.}~\bibnamefont {Volkmer}}, \ and\ \bibinfo {author}
  {\bibfnamefont {C.}~\bibnamefont {Seidel}},\ }\href@noop {} {\bibfield
  {journal} {\bibinfo  {journal} {Journal of biotechnology}\ }\textbf {\bibinfo
  {volume} {86}},\ \bibinfo {pages} {163} (\bibinfo {year} {2001})}\BibitemShut
  {NoStop}%
\bibitem [{\citenamefont {Orrit}(2009)}]{orrit2009langmuir}%
  \BibitemOpen
  \bibfield  {author} {\bibinfo {author} {\bibfnamefont {M.}~\bibnamefont
  {Orrit}},\ }\href@noop {} {\bibfield  {journal} {\bibinfo  {journal}
  {Colloids and Surfaces B: Biointerfaces}\ }\textbf {\bibinfo {volume} {74}},\
  \bibinfo {pages} {396} (\bibinfo {year} {2009})}\BibitemShut {NoStop}%
\bibitem [{\citenamefont {Debye}\ and\ \citenamefont
  {H{\"u}ckel}(1923)}]{debye1923theorie}%
  \BibitemOpen
  \bibfield  {author} {\bibinfo {author} {\bibfnamefont {P.~v.}\ \bibnamefont
  {Debye}}\ and\ \bibinfo {author} {\bibfnamefont {E.}~\bibnamefont
  {H{\"u}ckel}},\ }\href@noop {} {\bibfield  {journal} {\bibinfo  {journal}
  {phys. Z}\ }\textbf {\bibinfo {volume} {24}},\ \bibinfo {pages} {185}
  (\bibinfo {year} {1923})}\BibitemShut {NoStop}%
\bibitem [{\citenamefont {French}\ \emph {et~al.}(2010)\citenamefont {French},
  \citenamefont {Parsegian}, \citenamefont {Podgornik}, \citenamefont {Rajter},
  \citenamefont {Jagota}, \citenamefont {Luo}, \citenamefont {Asthagiri},
  \citenamefont {Chaudhury}, \citenamefont {Chiang}, \citenamefont {Granick}
  \emph {et~al.}}]{french2010long}%
  \BibitemOpen
  \bibfield  {author} {\bibinfo {author} {\bibfnamefont {R.~H.}\ \bibnamefont
  {French}}, \bibinfo {author} {\bibfnamefont {V.~A.}\ \bibnamefont
  {Parsegian}}, \bibinfo {author} {\bibfnamefont {R.}~\bibnamefont
  {Podgornik}}, \bibinfo {author} {\bibfnamefont {R.~F.}\ \bibnamefont
  {Rajter}}, \bibinfo {author} {\bibfnamefont {A.}~\bibnamefont {Jagota}},
  \bibinfo {author} {\bibfnamefont {J.}~\bibnamefont {Luo}}, \bibinfo {author}
  {\bibfnamefont {D.}~\bibnamefont {Asthagiri}}, \bibinfo {author}
  {\bibfnamefont {M.~K.}\ \bibnamefont {Chaudhury}}, \bibinfo {author}
  {\bibfnamefont {Y.-m.}\ \bibnamefont {Chiang}}, \bibinfo {author}
  {\bibfnamefont {S.}~\bibnamefont {Granick}},  \emph {et~al.},\ }\href@noop {}
  {\bibfield  {journal} {\bibinfo  {journal} {Reviews of Modern Physics}\
  }\textbf {\bibinfo {volume} {82}},\ \bibinfo {pages} {1900} (\bibinfo {year}
  {2010})}\BibitemShut {NoStop}%
\bibitem [{\citenamefont {Enderlein}\ \emph {et~al.}(1998)\citenamefont
  {Enderlein}, \citenamefont {Robbins}, \citenamefont {Ambrose},\ and\
  \citenamefont {Keller}}]{Joerg1998}%
  \BibitemOpen
  \bibfield  {author} {\bibinfo {author} {\bibfnamefont {J.}~\bibnamefont
  {Enderlein}}, \bibinfo {author} {\bibfnamefont {D.~L.}\ \bibnamefont
  {Robbins}}, \bibinfo {author} {\bibfnamefont {W.~P.}\ \bibnamefont
  {Ambrose}}, \ and\ \bibinfo {author} {\bibfnamefont {R.~A.}\ \bibnamefont
  {Keller}},\ }\href@noop {} {\bibfield  {journal} {\bibinfo  {journal} {The
  Journal of Physical Chemistry A}\ }\textbf {\bibinfo {volume} {102}},\
  \bibinfo {pages} {6089} (\bibinfo {year} {1998})}\BibitemShut {NoStop}%
\bibitem [{\citenamefont {Malek}\ and\ \citenamefont
  {Coppens}(2003)}]{malek2003knudsen}%
  \BibitemOpen
  \bibfield  {author} {\bibinfo {author} {\bibfnamefont {K.}~\bibnamefont
  {Malek}}\ and\ \bibinfo {author} {\bibfnamefont {M.-O.}\ \bibnamefont
  {Coppens}},\ }\href@noop {} {\bibfield  {journal} {\bibinfo  {journal} {The
  Journal of chemical physics}\ }\textbf {\bibinfo {volume} {119}},\ \bibinfo
  {pages} {2801} (\bibinfo {year} {2003})}\BibitemShut {NoStop}%
\bibitem [{\citenamefont {Li}\ \emph {et~al.}(2019)\citenamefont {Li},
  \citenamefont {Wang}, \citenamefont {Zhang},\ and\ \citenamefont
  {Qiao}}]{li2019diffusion}%
  \BibitemOpen
  \bibfield  {author} {\bibinfo {author} {\bibfnamefont {S.}~\bibnamefont
  {Li}}, \bibinfo {author} {\bibfnamefont {Y.}~\bibnamefont {Wang}}, \bibinfo
  {author} {\bibfnamefont {K.}~\bibnamefont {Zhang}}, \ and\ \bibinfo {author}
  {\bibfnamefont {C.}~\bibnamefont {Qiao}},\ }\href@noop {} {\bibfield
  {journal} {\bibinfo  {journal} {Industrial \& Engineering Chemistry
  Research}\ }\textbf {\bibinfo {volume} {58}},\ \bibinfo {pages} {21772}
  (\bibinfo {year} {2019})}\BibitemShut {NoStop}%
\bibitem [{\citenamefont {Wang}\ and\ \citenamefont
  {Schwartz}(2020)}]{wang2020non}%
  \BibitemOpen
  \bibfield  {author} {\bibinfo {author} {\bibfnamefont {D.}~\bibnamefont
  {Wang}}\ and\ \bibinfo {author} {\bibfnamefont {D.~K.}\ \bibnamefont
  {Schwartz}},\ }\href@noop {} {\bibfield  {journal} {\bibinfo  {journal} {The
  Journal of Physical Chemistry C}\ }\textbf {\bibinfo {volume} {124}},\
  \bibinfo {pages} {19880} (\bibinfo {year} {2020})}\BibitemShut {NoStop}%
\bibitem [{\citenamefont {Basch{\'e}}\ \emph {et~al.}(1992)\citenamefont
  {Basch{\'e}}, \citenamefont {Moerner}, \citenamefont {Orrit},\ and\
  \citenamefont {Talon}}]{basche1992photon}%
  \BibitemOpen
  \bibfield  {author} {\bibinfo {author} {\bibfnamefont {T.}~\bibnamefont
  {Basch{\'e}}}, \bibinfo {author} {\bibfnamefont {W.}~\bibnamefont {Moerner}},
  \bibinfo {author} {\bibfnamefont {M.}~\bibnamefont {Orrit}}, \ and\ \bibinfo
  {author} {\bibfnamefont {H.}~\bibnamefont {Talon}},\ }\href@noop {}
  {\bibfield  {journal} {\bibinfo  {journal} {Physical review letters}\
  }\textbf {\bibinfo {volume} {69}},\ \bibinfo {pages} {1516} (\bibinfo {year}
  {1992})}\BibitemShut {NoStop}%
\bibitem [{\citenamefont {Ghosh}\ \emph
  {et~al.}(2014{\natexlab{a}})\citenamefont {Ghosh}, \citenamefont {Chizhik},
  \citenamefont {Karedla}, \citenamefont {Dekaliuk}, \citenamefont {Gregor},
  \citenamefont {Schuhmann}, \citenamefont {Seibt}, \citenamefont {Bodensiek},
  \citenamefont {Schaap}, \citenamefont {Schulz} \emph {et~al.}}]{Ghosh2014}%
  \BibitemOpen
  \bibfield  {author} {\bibinfo {author} {\bibfnamefont {S.}~\bibnamefont
  {Ghosh}}, \bibinfo {author} {\bibfnamefont {A.~M.}\ \bibnamefont {Chizhik}},
  \bibinfo {author} {\bibfnamefont {N.}~\bibnamefont {Karedla}}, \bibinfo
  {author} {\bibfnamefont {M.~O.}\ \bibnamefont {Dekaliuk}}, \bibinfo {author}
  {\bibfnamefont {I.}~\bibnamefont {Gregor}}, \bibinfo {author} {\bibfnamefont
  {H.}~\bibnamefont {Schuhmann}}, \bibinfo {author} {\bibfnamefont
  {M.}~\bibnamefont {Seibt}}, \bibinfo {author} {\bibfnamefont
  {K.}~\bibnamefont {Bodensiek}}, \bibinfo {author} {\bibfnamefont {I.~A.}\
  \bibnamefont {Schaap}}, \bibinfo {author} {\bibfnamefont {O.}~\bibnamefont
  {Schulz}},  \emph {et~al.},\ }\href@noop {} {\bibfield  {journal} {\bibinfo
  {journal} {Nano letters}\ }\textbf {\bibinfo {volume} {14}},\ \bibinfo
  {pages} {5656} (\bibinfo {year} {2014}{\natexlab{a}})}\BibitemShut {NoStop}%
\bibitem [{\citenamefont {Piran}(2005)}]{piran2005physics}%
  \BibitemOpen
  \bibfield  {author} {\bibinfo {author} {\bibfnamefont {T.}~\bibnamefont
  {Piran}},\ }\href@noop {} {\bibfield  {journal} {\bibinfo  {journal} {Reviews
  of Modern Physics}\ }\textbf {\bibinfo {volume} {76}},\ \bibinfo {pages}
  {1143} (\bibinfo {year} {2005})}\BibitemShut {NoStop}%
\bibitem [{\citenamefont {Torella}\ \emph {et~al.}(2011)\citenamefont
  {Torella}, \citenamefont {Holden}, \citenamefont {Santoso}, \citenamefont
  {Hohlbein},\ and\ \citenamefont {Kapanidis}}]{torella2011identifying}%
  \BibitemOpen
  \bibfield  {author} {\bibinfo {author} {\bibfnamefont {J.~P.}\ \bibnamefont
  {Torella}}, \bibinfo {author} {\bibfnamefont {S.~J.}\ \bibnamefont {Holden}},
  \bibinfo {author} {\bibfnamefont {Y.}~\bibnamefont {Santoso}}, \bibinfo
  {author} {\bibfnamefont {J.}~\bibnamefont {Hohlbein}}, \ and\ \bibinfo
  {author} {\bibfnamefont {A.~N.}\ \bibnamefont {Kapanidis}},\ }\href@noop {}
  {\bibfield  {journal} {\bibinfo  {journal} {Biophysical journal}\ }\textbf
  {\bibinfo {volume} {100}},\ \bibinfo {pages} {1568} (\bibinfo {year}
  {2011})}\BibitemShut {NoStop}%
\bibitem [{\citenamefont {Chen}\ and\ \citenamefont
  {Dovichi}(1996)}]{chen1996single}%
  \BibitemOpen
  \bibfield  {author} {\bibinfo {author} {\bibfnamefont {D.}~\bibnamefont
  {Chen}}\ and\ \bibinfo {author} {\bibfnamefont {N.~J.}\ \bibnamefont
  {Dovichi}},\ }\href@noop {} {\bibfield  {journal} {\bibinfo  {journal}
  {Analytical Chemistry}\ }\textbf {\bibinfo {volume} {68}},\ \bibinfo {pages}
  {690} (\bibinfo {year} {1996})}\BibitemShut {NoStop}%
\bibitem [{\citenamefont {Dertinger}\ \emph
  {et~al.}(2007{\natexlab{a}})\citenamefont {Dertinger}, \citenamefont
  {Pacheco}, \citenamefont {von~der Hocht}, \citenamefont {Hartmann},
  \citenamefont {Gregor},\ and\ \citenamefont {Enderlein}}]{Joerg2007}%
  \BibitemOpen
  \bibfield  {author} {\bibinfo {author} {\bibfnamefont {T.}~\bibnamefont
  {Dertinger}}, \bibinfo {author} {\bibfnamefont {V.}~\bibnamefont {Pacheco}},
  \bibinfo {author} {\bibfnamefont {I.}~\bibnamefont {von~der Hocht}}, \bibinfo
  {author} {\bibfnamefont {R.}~\bibnamefont {Hartmann}}, \bibinfo {author}
  {\bibfnamefont {I.}~\bibnamefont {Gregor}}, \ and\ \bibinfo {author}
  {\bibfnamefont {J.}~\bibnamefont {Enderlein}},\ }\href@noop {} {\bibfield
  {journal} {\bibinfo  {journal} {ChemPhysChem}\ }\textbf {\bibinfo {volume}
  {8}},\ \bibinfo {pages} {433} (\bibinfo {year}
  {2007}{\natexlab{a}})}\BibitemShut {NoStop}%
\bibitem [{\citenamefont {Chiantia}\ \emph {et~al.}(2006)\citenamefont
  {Chiantia}, \citenamefont {Ries}, \citenamefont {Kahya},\ and\ \citenamefont
  {Schwille}}]{chiantia2006combined}%
  \BibitemOpen
  \bibfield  {author} {\bibinfo {author} {\bibfnamefont {S.}~\bibnamefont
  {Chiantia}}, \bibinfo {author} {\bibfnamefont {J.}~\bibnamefont {Ries}},
  \bibinfo {author} {\bibfnamefont {N.}~\bibnamefont {Kahya}}, \ and\ \bibinfo
  {author} {\bibfnamefont {P.}~\bibnamefont {Schwille}},\ }\href@noop {}
  {\bibfield  {journal} {\bibinfo  {journal} {ChemPhysChem}\ }\textbf {\bibinfo
  {volume} {7}},\ \bibinfo {pages} {2409} (\bibinfo {year} {2006})}\BibitemShut
  {NoStop}%
\bibitem [{\citenamefont {Dittrich}\ and\ \citenamefont
  {Schwille}(2002)}]{dittrich2002spatial}%
  \BibitemOpen
  \bibfield  {author} {\bibinfo {author} {\bibfnamefont {P.~S.}\ \bibnamefont
  {Dittrich}}\ and\ \bibinfo {author} {\bibfnamefont {P.}~\bibnamefont
  {Schwille}},\ }\href@noop {} {\bibfield  {journal} {\bibinfo  {journal}
  {Analytical chemistry}\ }\textbf {\bibinfo {volume} {74}},\ \bibinfo {pages}
  {4472} (\bibinfo {year} {2002})}\BibitemShut {NoStop}%
\bibitem [{\citenamefont {Nienhaus}\ \emph {et~al.}(2013)\citenamefont
  {Nienhaus}, \citenamefont {Maffre},\ and\ \citenamefont
  {Nienhaus}}]{nienhaus2013studying}%
  \BibitemOpen
  \bibfield  {author} {\bibinfo {author} {\bibfnamefont {G.~U.}\ \bibnamefont
  {Nienhaus}}, \bibinfo {author} {\bibfnamefont {P.}~\bibnamefont {Maffre}}, \
  and\ \bibinfo {author} {\bibfnamefont {K.}~\bibnamefont {Nienhaus}},\ }in\
  \href@noop {} {\emph {\bibinfo {booktitle} {Methods in enzymology}}},\ Vol.\
  \bibinfo {volume} {519}\ (\bibinfo  {publisher} {Elsevier},\ \bibinfo {year}
  {2013})\ pp.\ \bibinfo {pages} {115--137}\BibitemShut {NoStop}%
\bibitem [{\citenamefont {Feynman}\ and\ \citenamefont
  {Brown}(2005)}]{feynman2005feynman}%
  \BibitemOpen
  \bibfield  {author} {\bibinfo {author} {\bibfnamefont {R.~P.}\ \bibnamefont
  {Feynman}}\ and\ \bibinfo {author} {\bibfnamefont {L.~M.}\ \bibnamefont
  {Brown}},\ }\href@noop {} {\emph {\bibinfo {title} {Feynman's thesis: a new
  approach to quantum theory}}}\ (\bibinfo  {publisher} {World Scientific},\
  \bibinfo {year} {2005})\BibitemShut {NoStop}%
\bibitem [{\citenamefont {Enderlein}\ \emph {et~al.}(1997)\citenamefont
  {Enderlein}, \citenamefont {Robbins}, \citenamefont {Ambrose}, \citenamefont
  {Goodwin},\ and\ \citenamefont {Keller}}]{enderlein1997statistics}%
  \BibitemOpen
  \bibfield  {author} {\bibinfo {author} {\bibfnamefont {J.}~\bibnamefont
  {Enderlein}}, \bibinfo {author} {\bibfnamefont {D.~L.}\ \bibnamefont
  {Robbins}}, \bibinfo {author} {\bibfnamefont {W.~P.}\ \bibnamefont
  {Ambrose}}, \bibinfo {author} {\bibfnamefont {P.~M.}\ \bibnamefont
  {Goodwin}}, \ and\ \bibinfo {author} {\bibfnamefont {R.~A.}\ \bibnamefont
  {Keller}},\ }\href@noop {} {\bibfield  {journal} {\bibinfo  {journal}
  {Bioimaging}\ }\textbf {\bibinfo {volume} {5}},\ \bibinfo {pages} {88}
  (\bibinfo {year} {1997})}\BibitemShut {NoStop}%
\bibitem [{\citenamefont {Dertinger}\ \emph
  {et~al.}(2007{\natexlab{b}})\citenamefont {Dertinger}, \citenamefont
  {Pacheco}, \citenamefont {von~der Hocht}, \citenamefont {Hartmann},
  \citenamefont {Gregor},\ and\ \citenamefont {Enderlein}}]{dertinger2007two}%
  \BibitemOpen
  \bibfield  {author} {\bibinfo {author} {\bibfnamefont {T.}~\bibnamefont
  {Dertinger}}, \bibinfo {author} {\bibfnamefont {V.}~\bibnamefont {Pacheco}},
  \bibinfo {author} {\bibfnamefont {I.}~\bibnamefont {von~der Hocht}}, \bibinfo
  {author} {\bibfnamefont {R.}~\bibnamefont {Hartmann}}, \bibinfo {author}
  {\bibfnamefont {I.}~\bibnamefont {Gregor}}, \ and\ \bibinfo {author}
  {\bibfnamefont {J.}~\bibnamefont {Enderlein}},\ }\href@noop {} {\bibfield
  {journal} {\bibinfo  {journal} {ChemPhysChem}\ }\textbf {\bibinfo {volume}
  {8}},\ \bibinfo {pages} {433} (\bibinfo {year}
  {2007}{\natexlab{b}})}\BibitemShut {NoStop}%
\bibitem [{\citenamefont {Walder}\ \emph {et~al.}(2011)\citenamefont {Walder},
  \citenamefont {Nelson},\ and\ \citenamefont {Schwartz}}]{walder2011single}%
  \BibitemOpen
  \bibfield  {author} {\bibinfo {author} {\bibfnamefont {R.}~\bibnamefont
  {Walder}}, \bibinfo {author} {\bibfnamefont {N.}~\bibnamefont {Nelson}}, \
  and\ \bibinfo {author} {\bibfnamefont {D.~K.}\ \bibnamefont {Schwartz}},\
  }\href@noop {} {\bibfield  {journal} {\bibinfo  {journal} {Physical review
  letters}\ }\textbf {\bibinfo {volume} {107}},\ \bibinfo {pages} {156102}
  (\bibinfo {year} {2011})}\BibitemShut {NoStop}%
\bibitem [{\citenamefont {Ghosh}\ \emph
  {et~al.}(2014{\natexlab{b}})\citenamefont {Ghosh}, \citenamefont {Kumbhakar},
  \citenamefont {Platen}, \citenamefont {Gregor},\ and\ \citenamefont
  {Enderlein}}]{Ghosh2014-1}%
  \BibitemOpen
  \bibfield  {author} {\bibinfo {author} {\bibfnamefont {S.}~\bibnamefont
  {Ghosh}}, \bibinfo {author} {\bibfnamefont {M.}~\bibnamefont {Kumbhakar}},
  \bibinfo {author} {\bibfnamefont {M.}~\bibnamefont {Platen}}, \bibinfo
  {author} {\bibfnamefont {I.}~\bibnamefont {Gregor}}, \ and\ \bibinfo {author}
  {\bibfnamefont {J.}~\bibnamefont {Enderlein}},\ }in\ \href@noop {} {\emph
  {\bibinfo {booktitle} {Single Molecule Spectroscopy and Superresolution
  Imaging VII}}},\ Vol.\ \bibinfo {volume} {8950}\ (\bibinfo {organization}
  {International Society for Optics and Photonics},\ \bibinfo {year} {2014})\
  p.\ \bibinfo {pages} {895008}\BibitemShut {NoStop}%
\bibitem [{\citenamefont {Cheng}(2017)}]{Hao2017}%
  \BibitemOpen
  \bibfield  {author} {\bibinfo {author} {\bibfnamefont {H.}~\bibnamefont
  {Cheng}},\ }\emph {\bibinfo {title} {Probing Molecular Stoichiometry by
  Photon Antibunching and Nanofluidics Assisted Imaging in Solution}},\
  \href@noop {} {Ph.D. thesis},\ \bibinfo  {school}
  {Georg-August-Universit{\"a}t G{\"o}ttingen} (\bibinfo {year}
  {2017})\BibitemShut {NoStop}%
\bibitem [{\citenamefont {K{\"u}hn}\ and\ \citenamefont
  {Sandoghdar}(2006)}]{kuhn2006modification}%
  \BibitemOpen
  \bibfield  {author} {\bibinfo {author} {\bibfnamefont {S.}~\bibnamefont
  {K{\"u}hn}}\ and\ \bibinfo {author} {\bibfnamefont {V.}~\bibnamefont
  {Sandoghdar}},\ }\href@noop {} {\bibfield  {journal} {\bibinfo  {journal}
  {Applied Physics B}\ }\textbf {\bibinfo {volume} {84}},\ \bibinfo {pages}
  {211} (\bibinfo {year} {2006})}\BibitemShut {NoStop}%
\bibitem [{\citenamefont {K{\"u}hn}\ \emph {et~al.}(2008)\citenamefont
  {K{\"u}hn}, \citenamefont {Mori}, \citenamefont {Agio},\ and\ \citenamefont
  {Sandoghdar}}]{kuhn2008modification}%
  \BibitemOpen
  \bibfield  {author} {\bibinfo {author} {\bibfnamefont {S.}~\bibnamefont
  {K{\"u}hn}}, \bibinfo {author} {\bibfnamefont {G.}~\bibnamefont {Mori}},
  \bibinfo {author} {\bibfnamefont {M.}~\bibnamefont {Agio}}, \ and\ \bibinfo
  {author} {\bibfnamefont {V.}~\bibnamefont {Sandoghdar}},\ }\href@noop {}
  {\bibfield  {journal} {\bibinfo  {journal} {Molecular Physics}\ }\textbf
  {\bibinfo {volume} {106}},\ \bibinfo {pages} {893} (\bibinfo {year}
  {2008})}\BibitemShut {NoStop}%
\bibitem [{\citenamefont {Liaw}\ \emph {et~al.}(2009)\citenamefont {Liaw},
  \citenamefont {Chen}, \citenamefont {Chen},\ and\ \citenamefont
  {Kuo}}]{liaw2009purcell}%
  \BibitemOpen
  \bibfield  {author} {\bibinfo {author} {\bibfnamefont {J.-W.}\ \bibnamefont
  {Liaw}}, \bibinfo {author} {\bibfnamefont {J.-H.}\ \bibnamefont {Chen}},
  \bibinfo {author} {\bibfnamefont {C.-S.}\ \bibnamefont {Chen}}, \ and\
  \bibinfo {author} {\bibfnamefont {M.-K.}\ \bibnamefont {Kuo}},\ }\href@noop
  {} {\bibfield  {journal} {\bibinfo  {journal} {Optics express}\ }\textbf
  {\bibinfo {volume} {17}},\ \bibinfo {pages} {13532} (\bibinfo {year}
  {2009})}\BibitemShut {NoStop}%
\end{thebibliography}%

\clearpage
\newpage

\end{document}